\documentclass[sigplan,printcopyright,authorversion]{acmart}
\usepackage[utf8]{inputenc}
\usepackage[T1]{fontenc}
\usepackage{microtype}
\usepackage{siunitx}
\usepackage{multirow}
\usepackage{algorithm}
\usepackage{algorithmicx}
\usepackage{algpseudocode}
\usepackage[final]{listings}
\usepackage{enumitem}
\usepackage{flushend}
\usepackage{fancyhdr}

\title{Matching linear algebra and tensor code to specialized hardware accelerators}

\author{Pablo Antonio Martínez}
\email{pabloantonio.martinezs@um.es}
\orcid{0000-0002-4391-2451}
\affiliation{
\institution{University of Murcia}
\city{Murcia}
\country{Spain}
}

\author{Jackson Woodruff}
\email{J.C.woodruff@sms.ed.ac.uk}
\orcid{0000-0003-2650-9596}
\affiliation{
\institution{University of Edinburgh}
\city{Edinburgh}
\country{United Kingdom}
}

\author{Jordi Armengol-Estapé}
\email{jordi.armengol.estape@ed.ac.uk}
\orcid{0000-0001-8893-6185}
\affiliation{
\institution{University of Edinburgh}
\city{Edinburgh}
\country{United Kingdom}
}

\author{Gregorio Bernabé}
\email{gbernabe@um.es}
\orcid{0000-0002-7265-3508}
\affiliation{
\institution{University of Murcia}
\city{Murcia}
\country{Spain}
}

\author{José Manuel García}
\email{jmgarcia@um.es}
\orcid{0000-0002-6388-2835}
\affiliation{
\institution{University of Murcia}
\city{Murcia}
\country{Spain}
}

\author{Michael F.P. O’Boyle}
\email{mob@inf.ed.ac.uk}
\orcid{0000-0003-1619-5052}
\affiliation{
\institution{University of Edinburgh}
\city{Edinburgh}
\country{United Kingdom}
}

\renewcommand{\shortauthors}{P.A. Martínez, J. Woodruff, J. Armengol-Estapé, G. Bernabé, J.M. García, M.F.P. O’Boyle}

\setcopyright{rightsretained}
\acmPrice{15.00}
\acmDOI{10.1145/3578360.3580262}
\acmYear{2023}
\copyrightyear{2023}
\acmSubmissionID{cc23main-p40-p}
\acmISBN{979-8-4007-0088-0/23/02}
\acmConference[CC '23]{Proceedings of the 32nd ACM SIGPLAN International Conference on Compiler Construction}{February 25--26, 2023}{Montréal, QC, Canada}
\acmBooktitle{Proceedings of the 32nd ACM SIGPLAN International Conference on Compiler Construction (CC '23), February 25--26, 2023, Montréal, QC, Canada}

\definecolor{commentgreen}{RGB}{2,112,10}
\definecolor{eminence}{RGB}{108,48,130}
\definecolor{weborange}{RGB}{255,165,0}
\definecolor{background}{rgb}{0.95,0.95,0.93}

\lstdefinestyle{c}{
  emph={int,char,double,float,unsigned,void,bool},
  backgroundcolor=\color{background},
  basicstyle=\scriptsize\ttfamily,
  xleftmargin=0.7cm,
  frame=tlbr, 
  framesep=0.2cm, 
  framerule=0pt,
  captionpos=b,
  language=C,
  keywordstyle=\bf,
  showtabs=false,
  commentstyle=\color{commentgreen},
  keywordstyle=\color{eminence},
  stringstyle=\color{red},
  emphstyle={\color{blue}},
  classoffset=1, %
  otherkeywords={_mm256_load_ps, _mm256_broadcast_ss, _mm256_setzero_ps, _mm256_store_ps, _mm256_add_ps, _mm256_fmadd_ps},
  morekeywords={__m256},
  keywordstyle=\color{weborange},
  classoffset=0,
}

\lstdefinestyle{idl}{
  emph={int,char,double,float,unsigned,void,bool},
  backgroundcolor=\color{background},
  basicstyle=\scriptsize\ttfamily,
  xleftmargin=0.7cm,
  frame=tlbr, 
  framesep=0.2cm, 
  framerule=0pt,
  captionpos=b,
  language=C,
  keywordstyle=\bf,
  showtabs=false,
  commentstyle=\color{commentgreen},
  keywordstyle=\color{blue!80},
  stringstyle=\color{red},
  emphstyle={\color{blue}},
  classoffset=1, %
  otherkeywords={with, and, as},
  morekeywords={Constraint, End},
  keywordstyle=\color{eminence!60},
  classoffset=0,
}

\begin{document}

\begin{abstract}
Dedicated 
tensor accelerators 
demonstrate
the importance of linear algebra in modern 
applications.
Such accelerators have the potential for impressive performance gains, but require 
programmers to rewrite code using vendor APIs --- a 
barrier to wider
scale adoption.
Recent work
overcomes this by matching and replacing patterns
within code, but 
such %
approaches are fragile and fail to cope with the diversity of  
real-world codes.

We develop ATC, a compiler that uses program synthesis
to map regions of code to specific APIs.
The mapping space that ATC explores is combinatorially large,
requiring the development of program classification, dynamic analysis,
variable constraint generation and lexical distance matching techniques
to make it tractable.

We apply ATC to real-world tensor and linear algebra codes and evaluate them against four state-of-the-art approaches.
We accelerate between 2.6x and 7x more programs, leading to over an order of magnitude performance improvement.
\end{abstract}

\maketitle

\section{Introduction}

Linear algebra is a fundamental building block of many of today's critical applications; from weather modeling~\cite{Coiffier2011} to ubiquitous DNN~\cite{Deisenroth2020} workloads.
Its importance is reflected in the large number of accelerator libraries
and hardware devices devoted to fast linear algebra.
These range from specialized devices such as Google's TPU~\cite{tpu} to the tensor cores on NVIDIA~\cite{voltauarch}
among many others~\cite{IntelAIweb, Arm2020, Jouppi2021, Anderson2021, Fowers2019}.

While such devices promise significant performance for
an important class of
applications~\cite{Dally2020}, their uptake is limited by their
programmability~\cite{Domke2021}. Typically, these accelerators and libraries are accessed
via calls to specialized APIs, meaning existing code has to be rewritten. Given
the 
volume~\cite{Kalliamvakou2014} and variety~\cite{Livshits2015} of existing legacy code, such
rewriting is a %
significant undertaking~\cite{Dally2020}.

The combined importance of linear algebra acceleration and the difficulty
of rewriting legacy code to accelerators has led to recent %
work which attempts to automate the process.
These techniques search user code for matrix multiplications using constraints~\cite{idl,kernelfarer} or polyhedral analyses~\cite{Bhaskaracharya2020}
and replace regions of code with appropriate API calls or instructions.

However, as we show in Section~\ref{sec:cprograms}, these approaches are fragile.
 Constraints capture only a limited set of program patterns
and small variations in the user code defeat them. While they work
well on curated benchmarks, they perform poorly on real-world code~\cite{kernelfarer,facc},
defeated by function calls, optimized code and inline assembler.

Neural classification %
(e.g.~\cite{Cummins2021}) 
can effectively
detect code despite these challenges.
However, it does  not provide a path to acceleration, but requires
further steps.
These include  generating variable mappings and checking
for equivalence ~\cite{facc} which has shown promising results
for Fourier Transforms.
However, one of the key challenges in matching code to APIs is the cost of searching for user program variables that  map to API formal parameters. 
As the width of the API and complexity of the user program increase, this becomes combinatorially expensive.
As we show in Section
\ref{sec:complexity}
existing approaches \cite{facc} fail to scale to the
challenges that %
linear algebra APIs present.

We present ATC, a compiler that applies program synthesis to compile
general user-code to 
linear algebra
accelerators.  
We identify and solve  key %
challenges %

enabling the detect/synthesize paradigm 
to scale to the more complex APIs of linear algebra acceleration.
In addition, ATC employs 
a trained platform predictor 
to determine whether acceleration is profitable or not.

We applied our approach to 50 GitHub GEMM and 15 convolution projects  and discovered between 2.6 and 7x more linear operators
compared to KernelFaRer \cite{kernelfarer}, IDL \cite{idl}, Polly \cite{llvmpolly} or FACC\cite{facc}.  This resulted in  more than an order of magnitude performance improvement.

This paper makes the following contributions:
\begin{itemize}[nolistsep,noitemsep]
\item We present ATC, which maps matrix
multiplication and convolution programs to hardware accelerators,
up to 7x more frequently than existing techniques.
\item We introduce novel heuristics to reduce the mapping search space
by  four orders of magnitude.
\item We develop novel dynamic analyses to determine higher-level
information about variables, enabling synthesis without
costly whole-program analyses.
\end{itemize}

\section{Motivation} \label{sec:motivation}

\begin{figure}[ht]
\includegraphics[width=\linewidth]{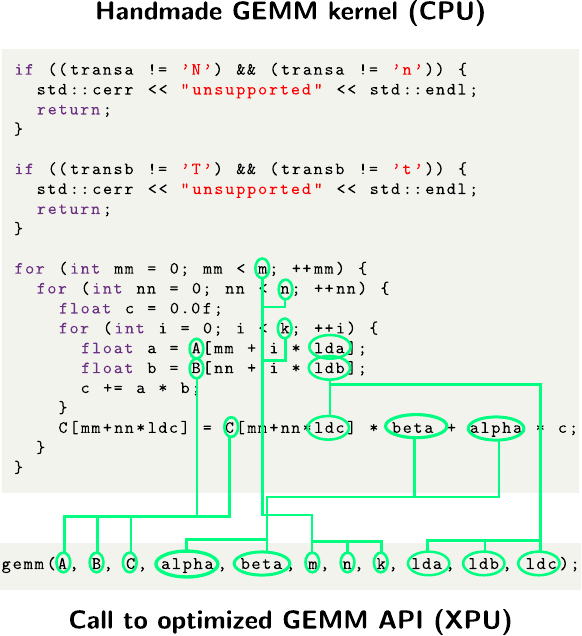}
\caption{Example application of API replacement. The above program is taken from the parboil benchmark~\cite{parboil}, a widely-used benchmark suite, which is transformed into a call to an optimized matrix-multiplication accelerator API.} \label{fig:example} 
\end{figure}

\subsection{Exisiting Match and replace}
\paragraph{IDL and KernelFaRer}  Both aim  to detect linear algebra operations in user programs and replace them with an appropriate accelerator library call. 
To illustrate this consider the code in Figure~\ref{fig:example}.
This shows a straight-forward matrix multiplication program fragment, from the parboil benchmark suite~\cite{parboil}. They aim to
detect this matrix-multiplication and replace it with a call to the library, shown at the bottom of the diagram.

To replace  code with an API call they have to both detect  the code performing  a  matrix multiplication and also determine which user program variables correspond to the arguments of the API call.
Both approaches are able to detect that this is a matrix multiplication, and
can determine the mapping between user variables and API parameters.

\subsection{Examples of complex GEMM programs}

\begin{figure}[t]
\centering
\includegraphics[width=\linewidth]{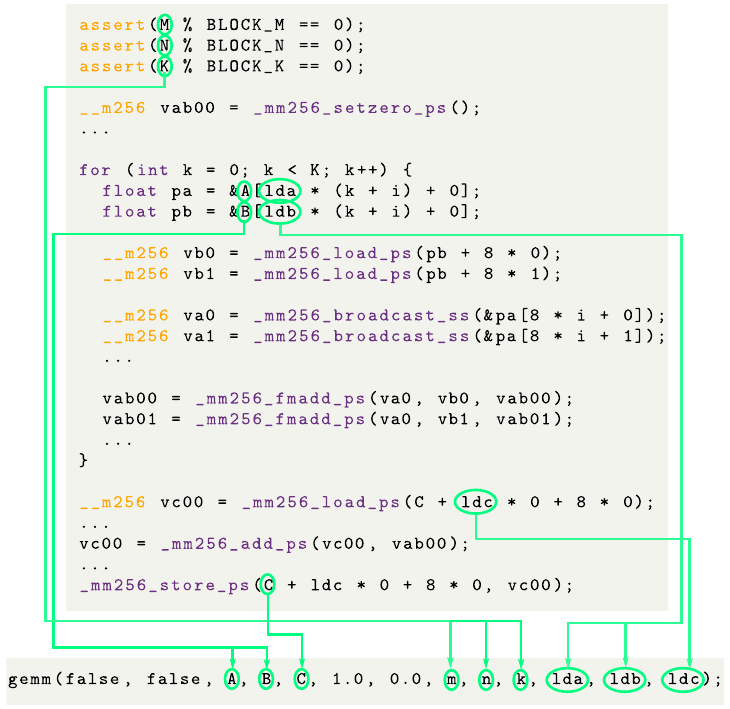} \label{fig:avx2}
\caption{GEMM code optimized for AVX2 found on GitHub consisting of 120 lines of hand-optimized intrinsics and how ATC matches the code to the accelerator API} \label{fig:avx2}
\end{figure}

Unfortunately, in practice, user code can be complex such that code structure or pattern-based  approaches
inevitably fail.

As an example, consider the code found on GitHub shown in Figure~\ref{fig:avx2}
which implements a matrix-multiplication algorithm (only a fragment of the 120 lines of user code are shown here).
The code structure is complex and difficult to understand as it makes extensive use 
 of inline assembler intrinsics which  defeats the code structure analysis approaches of  IDL and KernelFaRer, preventing acceleration.

\begin{figure*}[ht]
\includegraphics[width=0.9\linewidth]{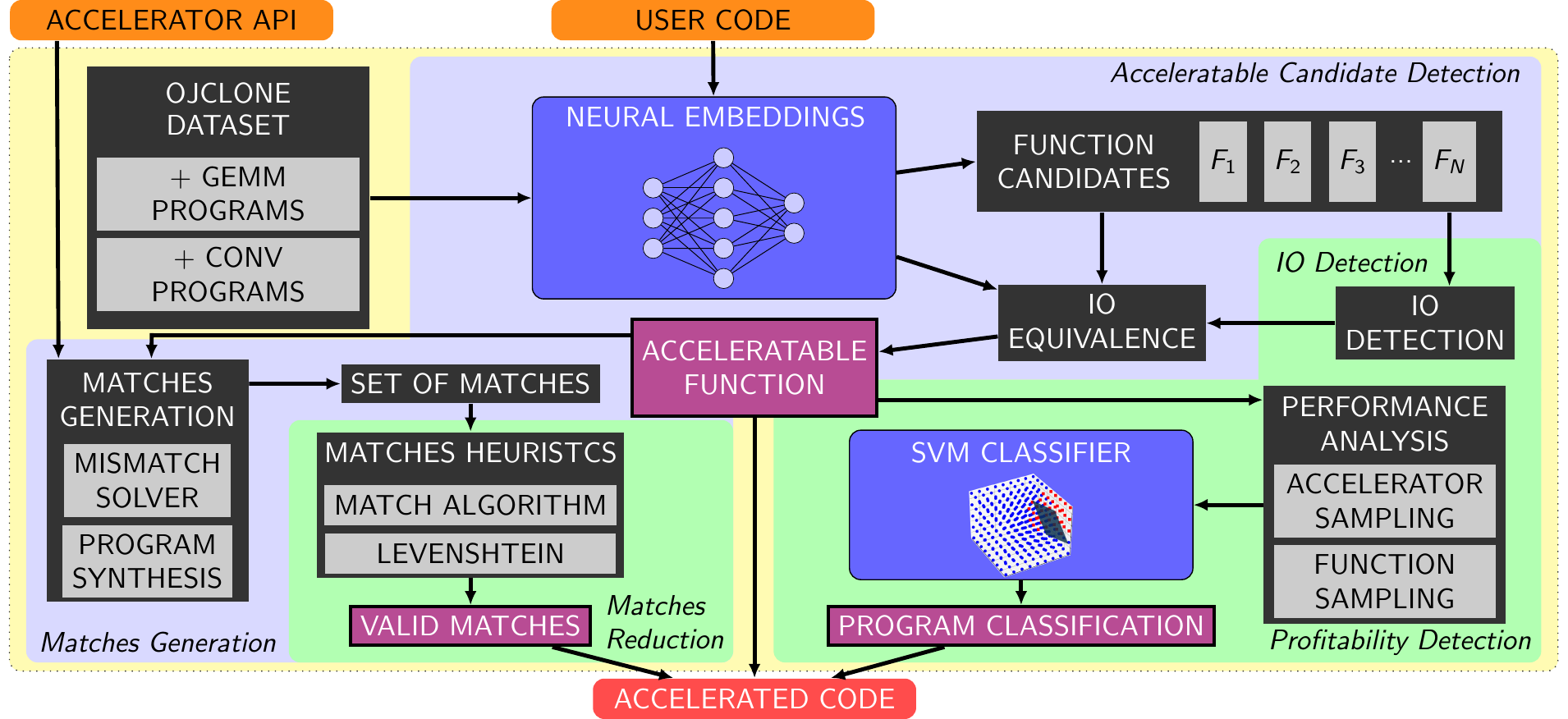}
\caption{ATC compiler architecture} \label{fig:architecture}
\end{figure*}
\subsection{Our approach - ATC}

Rather than relying on code structure to guide detection, ATC uses behavioral equivalence 
to determine if a section of code is a linear algebra operation.
Firstly, ATC uses neural program classification \cite{Cummins2021} to detect that the code in Figure \ref{fig:avx2} is probably a GEMM.
It then searches variable matches to determine the potential source and output arrays. As the search space is combinatorially large, we introduce scalable, algorithm-independent heuristics (which we discuss in Section~\ref{ReduingMatchingSpace}) that keep the number of mappings  manageable.
Next, ATC generates different input values for the arrays  and records the output.
After generating many randomized inputs, it observes that it has the equivalent behavior to the corresponding API and is able to replace the AVX2 code with the GEMM call at the bottom of Figure \ref{fig:avx2}.

\paragraph{Legality}
Now, IO behavioral equivalence is not proof
that a section of code is a particular linear algebra operation - similarly IDL and KernelFaRer do not prove equivalence. For proof, bounded model checking based on Kleene~\cite{Collie2022} can be deployed.
In practice, as demonstrated in our experimental section, IO equivalence gives no false positives. For further guarantees, we can ask for programmer sign-off or employ model checking.

\paragraph{Profitable}
Once we have detected and can  replace a section of code with an accelerator call, we need to determine if it is profitable to do. %
Due to hardware evolution, we do not use a hard-wired heuristic to determine profitability.
Instead, we learn, off-line,  a simple predictive model to determine if the target accelerator is faster than a CPU implementation.
The model is called at runtime, determining if offloading is worthwhile.

\paragraph{FACC}
Behavioral equivalence is also employed in FACC \cite{facc}.
Unfortunately, it is restricted to FFTs and one-dimensional arrays, and cannot detect the replacement in Figure~\ref{fig:example}.
Therefore, we extended FACC  to FACC* to consider GEMMs and multi-dimensional arrays.  This, however, exposes its weak variable binding model which is combinatorial in the number of user array variables and their dimensionality. 
Furthermore, it relies on program synthesis to determine the length of arrays,
which scales poorly to problems with many potential length parameters
for arrays such as GEMM\@. 

FACC also relies on brittle inter-procedural liveness analyses to determine
the liveness status of variables.  
This restricts it to running only at link time, rendering it invalid 
 for use in shared libraries. %
We will see in 
Section \ref{sec:evaluation} that the combination
of these issues results in excessively large search spaces.

\section{System overview} \label{sec:overview}
Figure~\ref{fig:architecture} gives a system flow overview of ATC.
We first detect regions of code that are likely to be linear algebraic operations using a neural program classifier. The classifier is trained ahead of time, based on programs that  are equivalent to the accelerator
and prior examples of linear algebra code. 

Once candidate code sections have been identified, we apply program   analysis to match user program variables with the particular API formal parameters. Given the combinatorially large  search space, we develop novel  
techniques to make the problem tractable.

For each candidate matching, we  generate multiple data inputs, execute the user code section and  record the output values. 
If the input/output pairs correspond to the input/output behavior of the accelerator API, we can say they are behavioral equivalent and candidates for replacement.

While candidate user code may be replaceable with a call to an accelerator API, it may not be profitable. 
Therefore, we employ a simple ML classifier, trained offline, and invoked at runtime to see if acceleration is appropriate for the user code for the runtime known array sizes.  

\subsection{ Neural Program Classification}

To detect potentially acceleratable parts of a program, we use prior work in  neural  program classification~\cite{Cummins2021}.
A network is trained with multiple instances of different program classes.
We use the OJClone dataset~\cite{ojclone}, which includes 105 classes of different programs, and add examples of the programs that we want to detect {\em e.g.} GEMMs and convolutions, gathered from benchmark suite repositories  other than GitHub.

At compile time a new  candidate program is divided into functions, which are presented to the neural classifier.
The classifier assigns each function in the program a probability of belonging to a certain class.  
We consider the most probable class, which in the case of a GEMM or convolution is then considered for variable matching and eventual code replacement
as described in the following sections.
Classification is fast ($\leq$ 1.5 sec) and has negligible impact on compilation time (see Section~\ref{sec:complexity}).

\section{Variable Matching} \label{sec:t1}
\label{VariableMatchingSection}

To check if a section of user code is behaviorally equivalent
to the API, we have to match up the user program variables with API
formal parameters. 
We first detect what variables are livein/liveout (Section~\ref{sec:liveinliveout}) and then the dimensions of arrays (Section~\ref{sec:dimensions}).

\subsection{Detecting livein and liveout variables} \label{sec:liveinliveout}
Detecting livein and liveout variables via standard static analysis is

straightforward for well-structured programs  
but fails for more diverse real-world codes,
which  may use assembly code or intrinsic functions.

ATC uses dynamic analysis to determine which variables are livein and liveouts inside a function.
In C, variables are passed by value so  non-pointers variables are always livein.
In the case of pointers (or arrays), we  generate random inputs with arbitrary sizes. 
If the values in memory change after executing the program, the array is considered liveout.

This allows us to detect which variables are livein or liveout, but not  both livein and liveout at the same time.
We generate a new random input  for liveout variables and re-execute the function.
If the output differs from the first execution, it is both livein and liveout.
We implement this algorithm as a just-in-time compiler pass in LLVM~\cite{llvm}.

\subsection{Detecting the dimensions of arrays} \label{sec:dimensions}
\begin{figure}[ht]
\includegraphics[width=\linewidth]{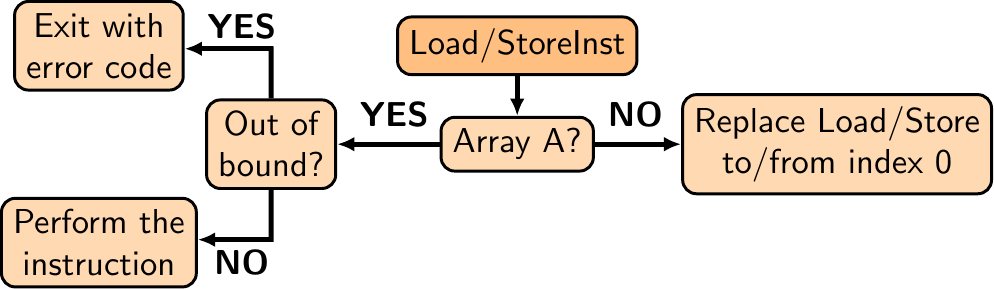}
\caption{Dimension detection algorithm overview for a target example array called A.} \label{fig:iodetection}
\end{figure}
Detecting arrays length enables offloading of appropriately-sized
regions of codes, so it is a critical step in ATC.
For some programs, lengths can be found  using static analysis (e.g.~\cite{Ravitch2009}), but this fails in more complex cases. 
We use runtime analysis to determine which program variables define 
array size using a modified form of runtime array bound checking. 
For each set of variables that could define an array's size (typically, from the argument list), we set such variables to a fixed value.
We then execute the user code that is modified to check runtime array accesses.  

\begin{algorithm}[t]
\caption{Dimensions detection algorithm} \label{alg:dims}
\begin{algorithmic}[1]
  \For{arr in function}
    \State $\Call{fakeLoadAndStoresExcept}{arr}$
    \State $\Call{replaceLoadAndStores}{arr}$ 
    \Repeat
      \State $c = \Call{getNextCombination}{arr}$
      \State $\Call{ffi\_call}{A, V}$
      \If{not failed}
        \State $found = True$
      \EndIf
    \Until{not found}
    \State Add $c$ to $C$
  \EndFor
  \State \Return $C$
\end{algorithmic}
\end{algorithm}

First, the compiler selects a target array to find its size.
Then, to generate the modified program, we tweak the load and store instructions in the user program, replacing them with custom function calls in the IR.
If a load or store does not access the array we are interested in, we modify it to load and store at a constant, safe location. 
If it does, the instruction is replaced with a function call that will check at runtime if the access is out of bounds. 
If so, the program exits with a custom error code. 
If not, we have found a valid array size.
The basic idea is depicted in Figure~\ref{fig:iodetection}.
This is used by our JIT analysis as shown in Algorithm~\ref{alg:dims} and implemented in LLVM\@.

This way, the compiler can assign different input sizes to a given array and check the exit code.
Therefore, the compiler iterates over all the possible dimensions combinations until one of the executions does not end with the custom error exit code.
That means that the program was completed without any illegal access to the target array, which indicates that it is the right dimension of the array.

\section{Reducing the matchings search space} \label{sec:t2}
\label{ReduingMatchingSpace}

To match code to APIs, the compiler generates
different candidates for the variable to formal parameter mappings
and then tests them using IO equivalence.
For small APIs, all mappings can be explored,
but the combinatorial cost makes it prohibitive for real-world accelerator APIs.
We develop techniques that reduce the mapping space by exploiting arrays information and human coding styles.

\subsection{Exploiting array information}
Using array dimensions (Section~\ref{sec:dimensions}),
we can reduce the number of possible matches that must
be checked, as assigning one array to another means that the
dimensions of each array must line up.

\subsubsection{Automatic matching algorithm} \label{sub:arrdim1}
\begin{algorithm}[t]
\caption{Automatic matching algorithm} \label{alg:algo}
\begin{algorithmic}[1]
\Function{dimsMatch}{$f1a, f2a, p, n$}
  \State $S = \emptyset$
  \State $idx \gets 0$
  \For{$args1$ in f1a}
    \State $args2$ = f2a[p[idx]]
    \State Add $\{args1, args2\}$ to $S$
    \State $idx \gets idx+1$
  \EndFor
  \State \Return \Call{Size}{S} = $n$
\EndFunction
\\
\Function{outMatch}{$f1o, f2o, p$} 
  \State idx = \Call{IndexOf}{$f2o$, 1}
  \State \Return \Call{IndexOf}{$p$, idx} = \Call{IndexOf}{$f1o$, 1}
\EndFunction
\\
\Function{findMatchings}{$f1a, f2a, f1o, f2o, n$}
  \State $B = \emptyset$
  \For{p in \Call{permutations}{$0$...$n$}}
    \If{\Call{dimsMatch}{f1a, f2a, p} \textbf{and} \\
       \ \ \ \ \ \ \ \ \ \ \ \ \ \ \ \ \Call{outMatch}{f1o, f2o, p}}
      \State Add $p$ to $B$
    \EndIf
  \EndFor
  \State \Return $B$
\EndFunction
\end{algorithmic}
\end{algorithm}
\begin{figure}[ht]
\includegraphics[width=\linewidth]{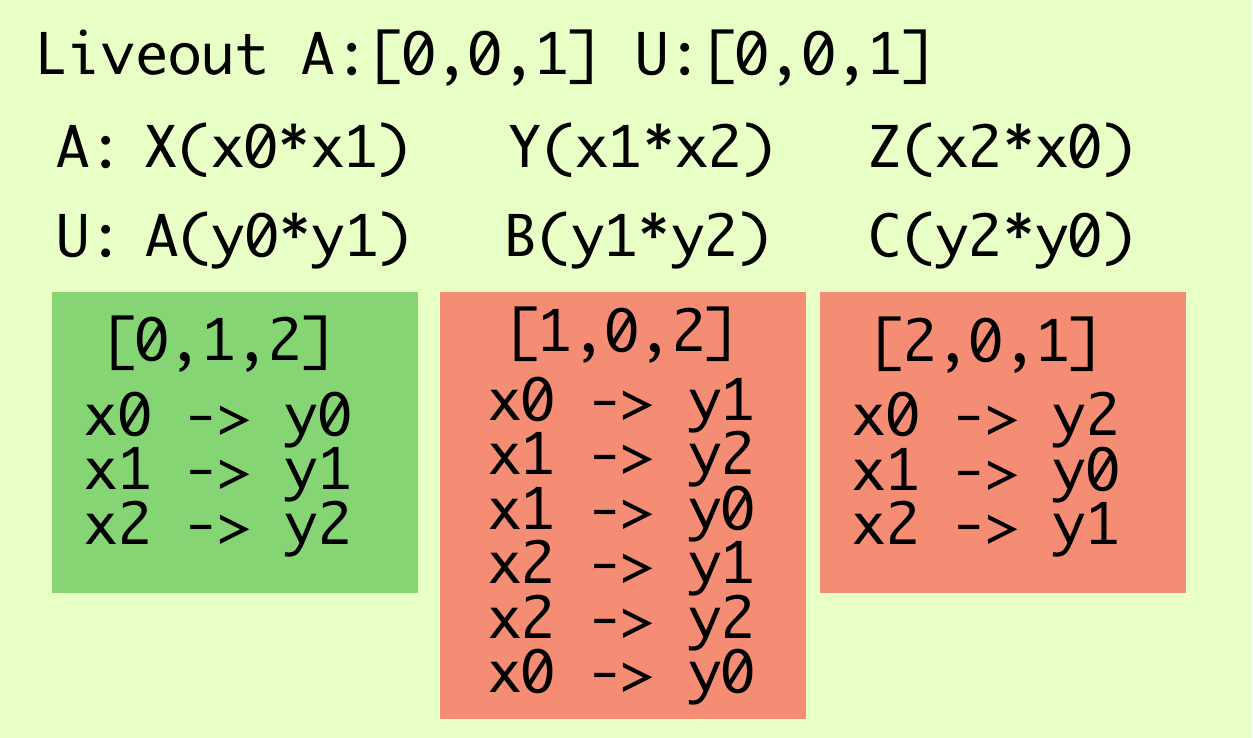}
\caption{Example application of the matching algorithm. The right match is found the algorithm automatically. Permutations in red means they are invalid, while the green permutation means valid.} \label{fig:examplealgo}
\end{figure}

We first generate all $n!$ permutations of  the $n$ array variables to $n$  parameters mapping.
We discard all permutations where variable livenesses do not match. Then for each candidate user array  and parameter array pair, we generate the constraints defining how their dimensions match. If we find contradictory constraints for any permutation, we discard it. The algorithm is shown in Algorithm~\ref{alg:algo}.

\subsubsection{Automatic Matching Algorithm: Example}

To illustrate this, Figure~\ref{fig:examplealgo} shows an example where we have two functions with three 2D arrays each.
First, the algorithm generates all the permutations between 0 and $n-1$ ($n=3$ in this example).
Then, for each permutation, it tries matching each variable in every array in the user code  with the corresponding variable in the array of the API (here we show only three of the six possible permutations).

In the first case (with the permutation $[0, 1, 2]$), the algorithm tries matching the array 
variables of the user  program $X, Y, Z$ with API parameters $A, B, C$ . We then examine each of the variables defining each of the corresponding arrays. Comparing $X$ and $A$ 
gives a match of $x0 \rightarrow y0$ and $x1 \rightarrow y1$.
For the second array variable $Y$ and API parameter $B$, we have  $x1 \rightarrow y1$ and $x2 \rightarrow y2$ and for the third variable pair $Z,C$ we have $x2 \rightarrow y2$ and $x0 \rightarrow y0$.
All of these are consistent with $n$=3 constraint, which
 satisfies the condition (\texttt{dimsMatch} in Algorithm~\ref{alg:algo}).
Liveout information is also satisfied
so this permutation is added as a potential mapping.

In the second permutation $[1,0,2]$,  where $X,Y,Z$ maps to $B,A,C$, the constraints are inconsistent {\em e.g.}
$x1\rightarrow y2$ and $x1 \rightarrow y0$ leading to 6 $\geq 3$,
so it is not a valid match.
In the third and last example, constraints are  equal to $n$, but the liveout arrays do not match.
Thus, the only valid match is the one found in the first permutation.

\subsection{Using argument names}
Programs are developed by humans, so we can assume that the functions that humans write follow common patterns.
We exploit this by analyzing the argument names of the API and the user program to find lexical similarities.

\begin{figure}[ht]
\centering
\begin{equation} \label{eq2}
lev(a,b) = \left\lbrace
\begin{array}{ll}
|a| & \textup{if }|b| = 0,\\
|b| & \textup{if }|a| = 0,\\
lev(tail(a), tail(b)) & \textup{if }a[0] = b[0],\\
1 + min \left\lbrace
\begin{array}{l}
lev(tail(a),b) \\
lev(a, tail(b)) \\
lev(tail(a),tail(b))
\end{array}
\right. & \textup{otherwise}
\end{array}
\right.
\end{equation}
\caption{Levenshtein recursive definition} \label{fig:editdistance}
\end{figure}

\begin{figure}[ht]
\includegraphics[width=\linewidth]{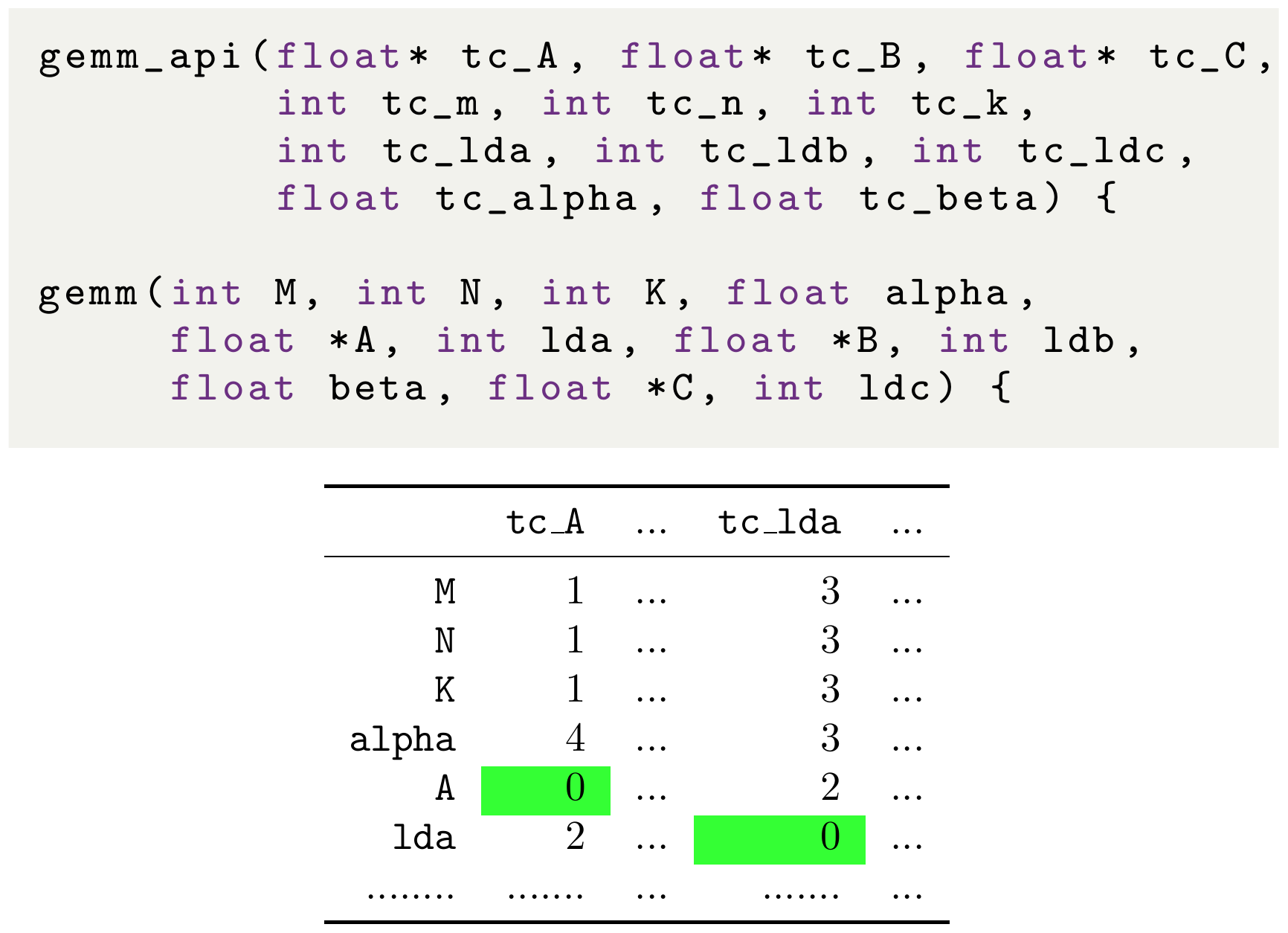}
\caption{Levenshtein distance calculation for the arguments of the tensor core API (above) and an example user program.} \label{fig:distancexample}
\end{figure}

To compare argument names, we use the Levenshtein distance~\cite{levenshtein} to compute the distance between each of the user programs and API arguments.
Figure~\ref{fig:editdistance} shows the definition of the Levenshtein distance, which calculation is based on the minimal number of modifications needed to transform one word into another, representing how close are those words.
After computing the distance, the compiler selects the combination that minimizes the Levenshtein distance.

Figure~\ref{fig:distancexample} shows an application example of the Levenshtein distance to a real case of GEMM matching.
For calculating the distance, we strip the API suffix (\texttt{tc\_}) and convert all names to lowercase.
Results show that the most probable mapping for \texttt{tc\_A} is \texttt{A} in the user code, and for \texttt{tc\_lda} is \texttt{lda}, which are the right matches.

\subsection{IO generation} \label{sec:iogen}

Once we have a candidate match we generate
random inputs of different sizes and test  for input-output (IO) equivalence. We use 30 inputs of varying sizes.
Although IO behavioral equivalence is not proof, we can increase the  number of tests for increased confidence. No existing technique such as IDL or KernelFaReR can prove that a matched piece of code is
provably equivalent to an API and therefore rely on user sign-off. 

\subsubsection{Behavioral Equivalence and the Limits of Verification}
ATC, like prior work on floating-point accelerators~\cite{facc}, uses
behavioral equivalence.  The downside of this strategy is that it requires
programmer sign-off to make any substitution.  However, due to the complexities
of verifying floating-point programs~\cite{facc}, verification of such liftings
are some way off.

In summary, the key challenges that all competing techniques face are:
\begin{itemize}[nolistsep,noitemsep]
\item Floating-point numbers often raise challenges in theorem provers
as they are challenging to reason about.
\item Floating-point functions may have different accuracies in different
input ranges, meaning that the obvious checks of correctness
(even within bounds) are difficult to apply.
\end{itemize}

The backend of ATC is not tied to using behavioral equivalence.  As we will
see, the use of such behavioral equivalence results in no false positives.
Further development of theorem prover technologies would mean that the weak
behavioral equivalence in ATC could easily be replaced with a theorem prover
guaranteeing correctness and enabling automatic transformations.

\section{Automatic profitability detection} \label{sec:t3}

We assume %
that user code runs faster when replaced by a platform-specific library.  The question is whether it is best to run on a CPU or accelerator version (XPU) of the library. This in turn depends on 
the input size, which is only known at runtime. 
We use a predictive model based on empirical data to
enable accurate predictions as platforms and libraries evolve by retraining the model.

\paragraph{SVM} We use the well-known
support vector machine (SVM) classifier with a polynomial kernel of degree 3 with gamma=1 and $C$=100.
We sample the CPU and the accelerator with a common
dataset of input sizes, which produces a dataset that is small enough to be processed in less than five minutes, but large enough to be highly  accurate.
Data is labeled with 0 or 1 meaning that the CPU or the XPU is faster. The model is then trained and deployed at runtime, when matrix sizes are known,
The training phase is done only once, at ``factory time'', and the resulting model when deployed has negligible ($ \leq 0.3 msec$) runtime overhead (see Section~\ref{sec:performance}).
\section{Setup} \label{sec:setup}
We evaluate GEMM and convolution acceleration on specialized platforms.
For GEMM, we used an Intel i7-11700 (CPU) with an NVIDIA Quadro RTX 5000 (tensor cores) (XPU).
For convolution, we used the Google Cloud Platform (GCP) services equipped with a TPUv3 with 8 TPU cores.
Compilation benchmarks in Section~\ref{sec:complexity} are executed in an AMD EPYC 7413.

The Intel/NVIDIA platform runs CentOS 8.3 with kernel 4.18.0. 
LLVM was downloaded from the official Git repository, using commit \texttt{329fda3}\@.
User codes were compiled using gcc 11.2.0 with \texttt{-O3 -march=native} flags.
We used cuBLAS 11.2 and MKL 2020.2.254 for compiling codes to the XPU and CPU, respectively.
For compiling convolution programs to the CPU, we used oneDNN v1.96.
The TPU system 
runs Debian 10 with kernel 4.19.0-14.

\subsection{User code} \label{sec:setupsw}
\begin{figure*}[t]
\centering
\begin{minipage}[t]{0.47\linewidth}
\resizebox{\linewidth}{!}{%
\begin{tabular}{|l|l|l|l|l|l|}
\hline
Algorithm                     & Code & LoC & Layout       & Sizes                  & Optimizations \\ \hline \hline 
\multirow{12}{*}{Naive}         & 1    & 22  & Column-major & Squared                & None          \\ \cline{2-6} 
                                & 2    & 127 & Both         & Any                    & None          \\ \cline{2-6} 
                                & 3    & 18  & Row-major    & Any                    & None          \\ \cline{2-6} 
                                & 4    & 41  & Column-major & Squared                & None          \\ \cline{2-6} 
                                & 5    & 11   & Row-major    & Any                    & None          \\ \cline{2-6} 
                                & 6    & 11   & Row-major    & Any                    & None          \\ \cline{2-6} 
                                & 7    & 30  & Row-major    & Any                    & None          \\ \cline{2-6} 
                                & 8    & 18  & Column-major & Any                    & None          \\ \cline{2-6} 
                                & 9    & 40  & Column-major & Any                    & None          \\ \cline{2-6} 
                                & 10   & 39  & Column-major & Any                    & None          \\ \cline{2-6} 
                                & 11   & 43  & Row-major    & Any                    & None          \\ \cline{2-6} 
                                & 12   & 11   & Row-major    & Squared                & None          \\ \hline
\multirow{5}{*}{\begin{tabular}[c]{@{}l@{}}Naive \\ parallel\end{tabular}} & 13   & 39  & Row-major    & Squared                & OpenMP        \\ \cline{2-6} 
                                & 14   & 28  & Column-major & Squared                & OpenMP        \\ \cline{2-6} 
                                & 15   & 164 & Row-major    & Any                    & OpenMP        \\ \cline{2-6} 
                                & 16   & 22  & Row-major    & Multiple of nthreads   & C++ threads   \\ \cline{2-6} 
                                & 17   & 107  & Row-major    & Squared                & C++ threads   \\ \hline
\multirow{4}{*}{Unrolled}       & 18   & 57  & Row-major    & Any                    & None          \\ \cline{2-6} 
                                & 19   & 50  & Row-major    & Any                    & None          \\ \cline{2-6} 
                                & 20   & 63  & Row-major    & Squared                & OpenMP        \\ \cline{2-6} 
                                & 21   & 38  & Row-major    & Squared, multiple of bs & None          \\ \hline
\multirow{4}{*}{Kernel Calls}   & 22   & 46  & Column-major & Any                    & None          \\ \cline{2-6} 
                                & 23   & 115 & Column-major & Any                    & OpenMP        \\ \cline{2-6} 
                                & 24   & 61  & Column-major & Any                    & None          \\ \cline{2-6} 
                                & 25   & 105 & Column-major & Any                    & Unrolled      \\ \hline
\end{tabular}
}
\end{minipage}\qquad
\begin{minipage}[t]{0.47\linewidth}
\resizebox{\linewidth}{!}{%
\begin{tabular}{|l|l|l|l|l|l|}
\hline
Algorithm                    & Code & LoC & Layout       & Sizes                  & Optimizations \\ \hline \hline 
Kernel Calls                 & 26 & 164 & Column-major & Any                 & Unrolled          \\ \hline
\multirow{9}{*}{Blocked}     & 27 & 104  & Row-major    & Any                 & Block             \\ \cline{2-6} 
                             & 28 & 30  & Row-major    & Squared             & OpenMP            \\ \cline{2-6} 
                             & 29 & 52  & Column-major & Any                 & None              \\ \cline{2-6} 
                             & 30 & 35  & Row-major    & Squared             & None              \\ \cline{2-6} 
                             & 31 & 38  & Column-major & Squared             & None              \\ \cline{2-6} 
                             & 32 & 42  & Row-major    & Multiple of bs      & Unrolled          \\ \cline{2-6} 
                             & 33 & 49  & Row-major    & Squared             & None              \\ \cline{2-6} 
                             & 34 & 18  & Row-major    & Squared             & None              \\ \cline{2-6} 
                             & 35 & 21  & Row-major    & Squared             & None              \\ \hline
\multirow{2}{*}{Goto}        & 36 & 247 & Column-major & Squared             & Intrinsics (SSE)  \\ \cline{2-6} 
                             & 37 & 89  & Row-major    & Squared             & None              \\ \hline
\multirow{3}{*}{Strassen}    & 38 & 210 & Row-major    & Squared             & None              \\ \cline{2-6} 
                             & 39 & 315 & Row-major    & Squared, power of 2 & None              \\ \cline{2-6} 
                             & 40 & 162 & Row-major    & Squared             & None              \\ \hline
\multirow{10}{*}{Intrinsics} & 41 & 102  & Row-major    & Squared             & Intrinsics (AVX2) \\ \cline{2-6} 
                             & 42 & 91  & Row-major    & Multiple of 8       & Intrinsics (AVX2) \\ \cline{2-6} 
                             & 43 & 82  & Row-major    & Multiple of 8       & Intrinsics (AVX2) \\ \cline{2-6} 
                             & 44 & 58  & Row-major    & Any                 & Intrinsics (SSE)  \\ \cline{2-6} 
                             & 45 & 112  & Row-major    & Multiple of bs      & Intrinsics (AVX2) \\ \cline{2-6} 
                             & 46 & 136 & Row-major    & Multiple of bs      & Intrinsics (AVX2) \\ \cline{2-6} 
                             & 47 & 120 & Row-major    & Any                 & Intrinsics (AVX2) \\ \cline{2-6} 
                             & 48 & 143 & Row-major    & Multiple of bs      & Intrinsics (AVX2) \\ \cline{2-6} 
                             & 49 & 57  & Row-major    & Multiple of bs      & Intrinsics (AVX2) \\ \cline{2-6} 
                             & 50 & 60  & Row-major    & Any                 & Intrinsics (SSE)  \\ \hline
\end{tabular}
}
\end{minipage}
\vspace*{-0.35cm}
\caption{List of GEMM codes} \label{gemmcodes}
\end{figure*}

\begin{table}[t]
\resizebox{1\linewidth}{!}{%
\begin{tabular}{|l|l|l|l|l|l|l|}
\hline
Algorithm                                                               & Code & LoC & Nº Args & Optimizations       & Constraints & C struct? \\ \hline \hline
\multirow{9}{*}{Direct}                                                 & 1    & 35  & 12      & None                & None        & No        \\ \cline{2-7}
                                                                        & 2    & 36  & 10      & OpenMP              & FW = FH = 3 & No        \\ \cline{2-7}
                                                                        & 3    & 34  & 8       & OpenMP              & FW = FH = 3 & No        \\ \cline{2-7}
                                                                        & 4    & 43  & 11      & None                & FW = FH = 3 & No        \\ \cline{2-7}
                                                                        & 5    & 39  & 8       & OpenMP              & FW = FH = 3 & No        \\ \cline{2-7}
                                                                        & 6    & 76  & 16      & None                & N = 1       & No        \\ \cline{2-7}
                                                                        & 7    & 209 & 18      & Vectorized          & N = 1       & Yes       \\ \cline{2-7}
                                                                        & 8    & 102 & 12      & None                & None        & No        \\ \cline{2-7}
                                                                        & 9    & 42  & 16      & None                & None        & No        \\ \hline
\multirow{3}{*}{\begin{tabular}[c]{@{}l@{}}im2col+\\ gemm\end{tabular}} & 10   & 189 & 15      & None                & N = 1       & Yes       \\ \cline{2-7}
                                                                        & 11   & 286 & 15      & BLAS                & N = 1       & Yes       \\ \cline{2-7}
                                                                        & 12   & 179 & 17      & BLAS                & FW = FH     & Yes       \\ \hline
\multirow{3}{*}{Winograd}                                               & 13   & 687 & 17      & Intrinsics + OpenMP & FW = FH = 3 & No        \\ \cline{2-7}
                                                                        & 14   & 254 & 12      & None                & N = 1       & Yes       \\ \cline{2-7}
                                                                        & 15   & 782 & 12      & Intrinsics + OpenMP & FW = FH = 3 & No        \\ \hline
\end{tabular}
}
\caption{List of convolution codes} \label{convcodes}
\end{table}

We explored GitHub looking for C and C++ GEMM codes, analyzing more than 400 programs from which we selected 50 programs.
We discarded the rest of them because of wrong implementations, compilation errors or duplicated code.
The final list of programs is shown in Table~\ref{gemmcodes}.
We categorize the codes as follows:
{\em Naive:} naive implementations with the traditional 3-loop structure;
{\em Naive Parallel:} as Naive  but with simple outer loop parallelization;
{\em Unrolled:} naive implementation with unrolled loops;
{\em Kernel Calls:} implementations that divide the loops into different function calls;
{\em Blocked:} tiled implementations;
{\em Goto:} implementations of the Goto algorithm~\cite{goto};
{\em Strassen:} implementations of the Strassen algorithm~\cite{strassen};
{\em Intrinsics:} implementations using Intel intrinsics.

In addition, we selected 50 non-GEMM projects to check whether any of the approaches gave false positives.

\paragraph{Convolutions} We explored GitHub looking for C and C++ 4D convolution implementations.
We analyzed around 50 programs from which we a selected list of 15 programs based on the same methodology used for selecting GEMMs.
The list of convolution programs is shown in Table~\ref{convcodes}.
We have included codes from the most relevant convolution implementations:
{\em Direct:} the direct convolution algorithm;
{\em im2col+gemm:}  an algorithm that casts the input as matrices (im2col) and later uses a GEMM, as in Caffe~\cite{caffe};
{\em Winograd:} the Winograd algorithm.

\subsection{Methods}
We evaluate our approach against 4 well known schemes:\\
{\bf IDL:}
Idioms are described using an idiom description language \cite{idl}, which is translated into a set of constraints over LLVM IR.\\
{\bf KernelFaRer:}
Uses different pattern matching to detect specific code constructs, matching specific matrix-multiplication structures \cite{kernelfarer}.\\
{\bf Polly:}
Detects static control parts (SCoPs) in the code using the polyhedral model \cite{llvmpolly}.
It does not replace the code with a call to an optimized library.\\
{\bf FACC*:}
FACC uses neural embeddings and behavioral synthesis to detect candidates for acceleration~\cite{facc}. It is limited to 1D arrays 
so we developed an extended version, FACC*, which supports multi-dimensional arrays. \\

\section{Results} \label{sec:evaluation}

\subsection{Detection} \label{sec:cprograms}

\begin{figure*}[t]
\includegraphics[width=\linewidth]{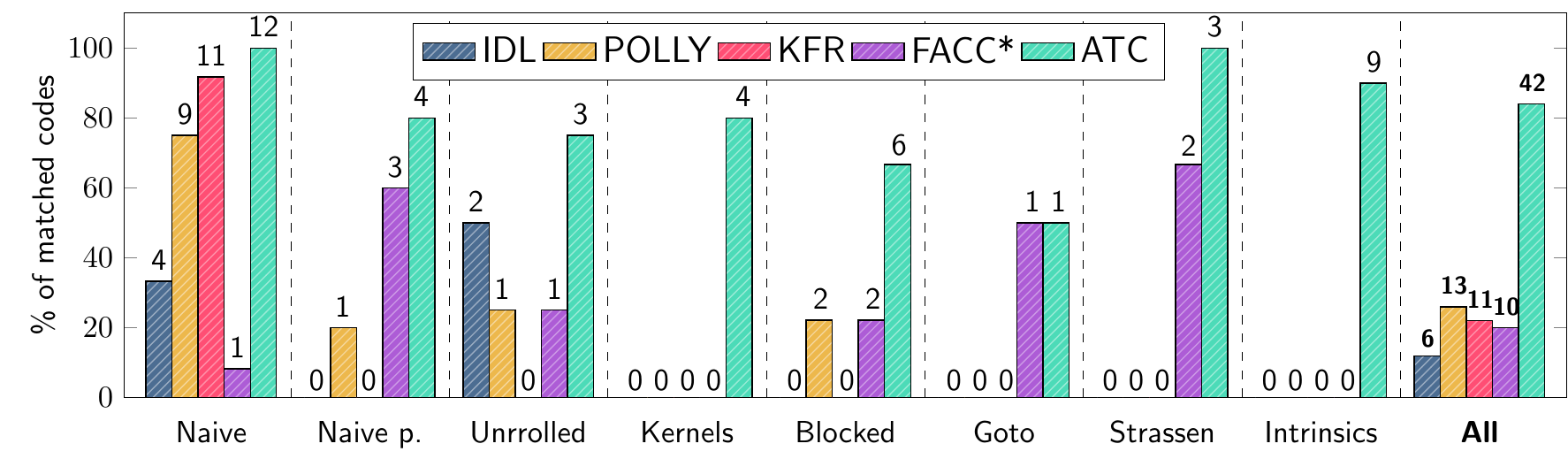}
\caption{Percentage of matched GEMM codes by different techniques.}
\label{fig:matched}
\end{figure*}

\begin{figure}[t]
\includegraphics[width=\linewidth]{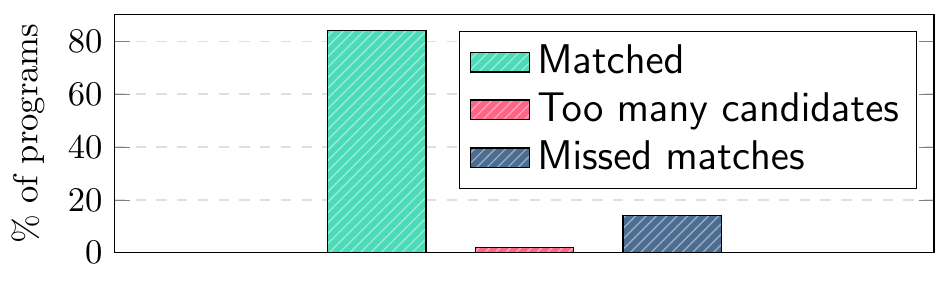}
\caption{Percentage of matched GEMM codes by ATC divided by failure reason.} \label{fig:atcsuccess}
\end{figure}

Figure~\ref{fig:matched} shows the percentage of GEMM programs matched by each technique across each  of  8 categories listed in Table~\ref{gemmcodes}.

\paragraph{IDL} The constraint based scheme~\cite{idl} only matches 6 out of 50 cases. 
These programs are largely naive implementations of GEMM, with a simple  loop structure.
It is able to manage 2 programs containing unrolled loops but fails on anything more complex.
Matching more diverse cases would require writing a new IDL constraint description for each sub-class.

\paragraph{KernelFaRer} This code matching approach~\cite{kernelfarer} is more successful, matching
11 GEMMs due  to a more robust pattern matcher. For straightforward  sequential 
implementations, it is able to match all but one of the cases.
However, any code variation, including loop unrolling, defeats it.

\paragraph{Polly}
Although it does not match and replace GEMMs, it can detect SCoPs which may be candidates for replacement
with appropriate API calls. It is less successful than KernelFaRer in detecting naive implementations but is more robust across other more complex categories including one parallel and unrolled versions and 2 blocked cases.
It slightly outperforms KernelFaRer, matching 13 vs. 11 out of 50 cases.

\paragraph{FACC*}  
Unlike the other approaches, FACC* performed poorly on naive implementations, but better on others.
Here, the size of  the mapping search space is the limiting factor.  
It  was able to find  10 cases in the available time 
(timeout $\leq$ 10 mins). We examine the reasons for this in Section~\ref{sec:complexity}.

\paragraph{ATC} Our approach  is significantly more robust across all categories,
matching 42 out of 50 cases. 
It is able to detect all naive implementations and the majority  within each other category.  
It detects more naive parallel implementations, unrolled and blocked programs than Polly and is the only technique to detect GEMMs in codes containing kernel calls and intrinsic instructions.

\subsubsection{Accuracy}

Figure~\ref{fig:atcsuccess} provides a summary of ATC's success and failure by 
type. In 8 cases ATC failed to detect that the program contained a GEMM.  In one
case, program 23, this is due to
there being too many candidate matches, 280 which is above our timeout threshold of 100 candidates. 
The remaining cases are due to overly aggressive search pruning, missing a legal match.
Improved search heuristics are likely to improve program coverage.

\paragraph{False positives} None of the methods classified any of the 50  non-GEMMs as a GEMM\@.  Across all methods, there were no false positives.

\subsection{Performance}
\label{sec:performance}
\begin{figure*}[t]
\includegraphics[width=0.9\linewidth]{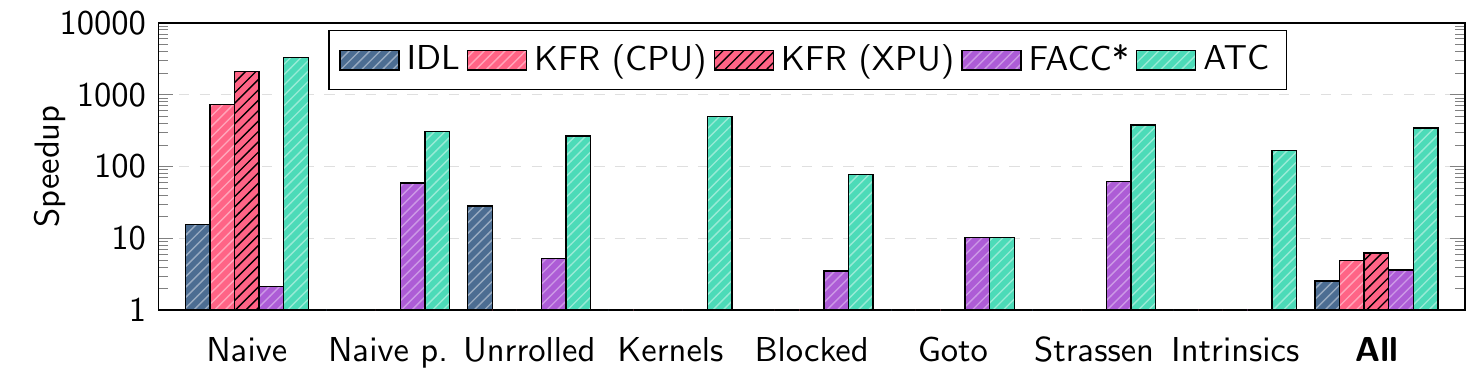}
\caption{Geometric mean speedup obtained by IDL, KernelFaRer, FACC* and ATC in GEMM programs with 
$n=8192$.}
\label{fig:speedups}
\end{figure*}

The performance of each approach 
is shown in Figure
\ref{fig:speedups}.
Polly is not included here as although it can detect SCoPs, it does not explicitly identify them as GEMMs for API replacement.
We show two bars for KernelFaRer, which correspond to the strategy of GEMM code with an optimized CPU implementation as described in~\cite{kernelfarer} and KFR (XPU) which is our extension, replacing the CPU library with the optimized XPU implementation. 
IDL and FACC* directly target the accelerator, while ATC chooses the CPU or accelerator based on its SVM platform predictor.  This runtime prediction cost is negligible $\leq 0.3msec$ and included in Figure~\ref{fig:speedups}.

What is immediately clear is that detecting more GEMMs leads to better overall speedup. In the Naive category, KFR and ATC are both able to achieve good performance, with a speedup of 726x and 1031x, respectively. The gap is narrowed when using KFR (XPU).  However, KFR is unable to detect GEMMs in any other category leading to just a 6.2x speedup overall while ATC achieves 344.0x. Unsurprisingly, there is more performance available on naive sequential implementations than in those cases where the programmer has spent effort in optimizing the program.%

\subsection{Candidate search complexity and compile time}
\label{sec:complexity}
 
\begin{figure*}[t]
\includegraphics[width=0.95\linewidth]{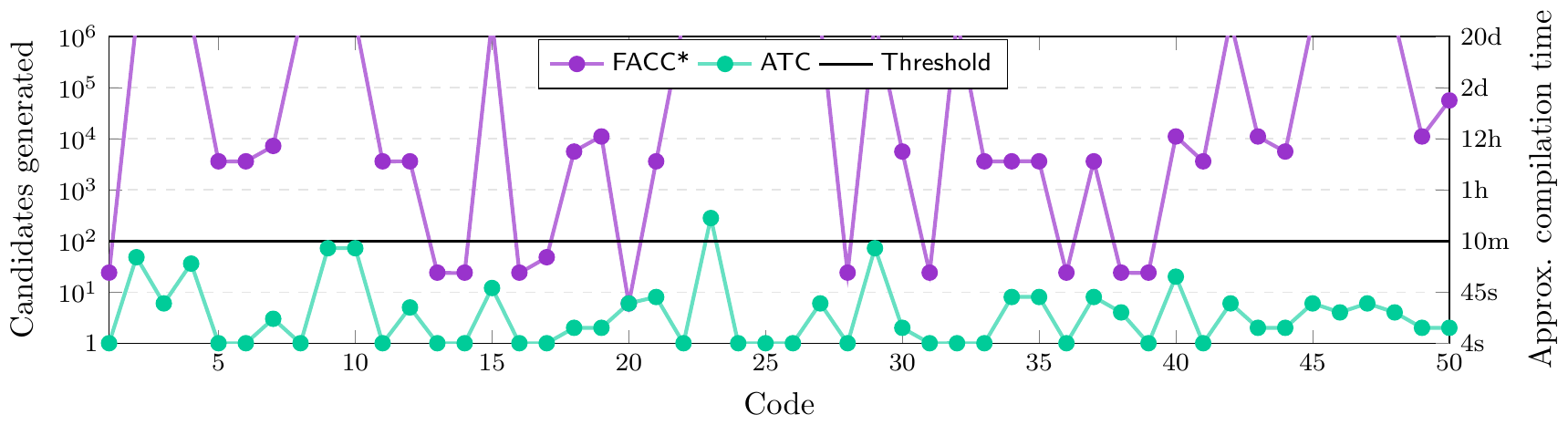}
\caption{Comparison of the number of candidates generated for matching GEMM codes: FACC* vs our approach.} \label{fig:numcand}
\end{figure*}

One of the key challenges in matching code to APIs is searching for program variables that  map to API formal parameters. 
As the width of the API and complexity of the user program increase, this becomes combinatorially expensive. 
Figure~\ref{fig:numcand} evaluates FACC* naive matching of variables and our approach based on the Levenshtein distance.
Naive matching varies considerably from just 4 candidates  to over 1 million. 
Our approach greatly reduces the number of candidates for the majority of the programs.
There is one special case, code 23, where we reduce the number of candidates, but it is still too high. %

\begin{figure}[t]
\includegraphics[width=\linewidth]{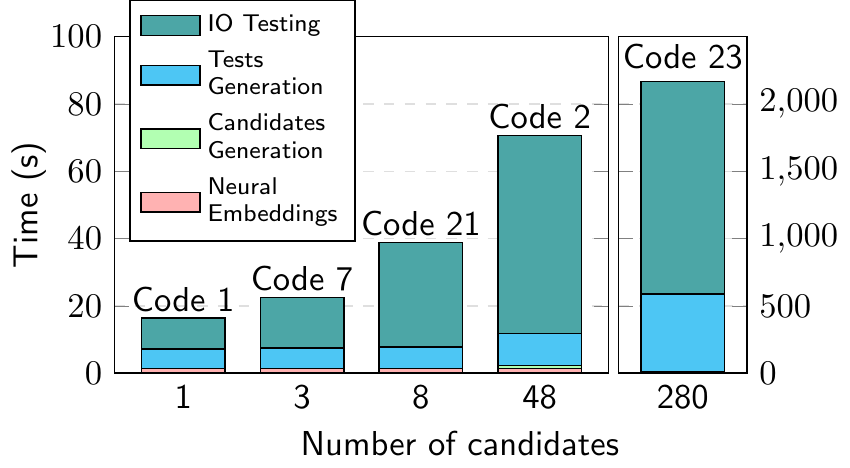}
\caption{Compilation time for different number of candidates.} \label{fig:comptime}
\end{figure}

Figure~\ref{fig:comptime} shows the compilation time of ATC. 
The initial neural classifier has a negligible constant execution time of 1.3 seconds, while the other phases'  compilation time grows with the number of candidates.

As the number of candidates begins to increase 
compilation time becomes prohibitively expensive.
Code 23 has 280 candidates which would take 35 mins more to evaluate.
We limit the number of candidates considered to 100 which corresponds to a timeout of $\leq 10$  minutes.

\subsection{Profitability accuracy}
To measure the accuracy of the SVM platform predictor,  we built a model offline and 
tested it on unseen data values. 

\begin{table}[t]
\resizebox{\linewidth}{!}{%
\begin{tabular}{lllllll}
\toprule
\multirow{2}{*}{\begin{tabular}[c]{@{}l@{}}\\[-16pt]Parameter\\[-0pt] Value\\[-0pt] (mnk)\end{tabular}} & \multicolumn{5}{l}{\begin{tabular}[c]{@{}l@{}} \\m\end{tabular}}                                                                                                         & \multirow{2}{*}{\begin{tabular}[c]{@{}l@{}}Global \\ Accuracy\end{tabular}} \\ \cmidrule{2-6}
                                                                                & \multicolumn{1}{l}{2000}  & \multicolumn{1}{l}{4000}   & \multicolumn{1}{l}{6000}  & \multicolumn{1}{l}{8000}   & 10000 &                                                                             \\ \midrule
111                                                                             & \multicolumn{1}{l}{100\%} & \multicolumn{1}{l}{100\%}  & \multicolumn{1}{l}{100\%} & \multicolumn{1}{l}{70.0\%} & 100\% & 93.8\%                                                                      \\
123                                                                             & \multicolumn{1}{l}{100\%} & \multicolumn{1}{l}{78.9\%} & \multicolumn{1}{l}{100\%} & \multicolumn{1}{l}{100\%}  & 100\% & 95.9\%                                                                      \\
312                                                                             & \multicolumn{1}{l}{100\%} & \multicolumn{1}{l}{84.3\%} & \multicolumn{1}{l}{100\%} & \multicolumn{1}{l}{100\%}  & 100\% & 96.9\%                                                                      \\
136                                                                             & \multicolumn{1}{l}{100\%} & \multicolumn{1}{l}{89.5\%} & \multicolumn{1}{l}{100\%} & \multicolumn{1}{l}{100\%}  & 100\% & 97.9\%                                                                      \\
\bottomrule
\end{tabular}
}
\caption{SVM accuracy for different sizes. 
111 means m = 1 $\times$ m, n = 1 $\times$ m, k = 1 $\times$ m. 
123 means m = 1 $\times$ m, n = 2 $\times$ m, k = 3 $\times$ m etc} \label{tab:svm}
\end{table}

Table~\ref{tab:svm} summarizes the SVM accuracy with different input sizes and shapes.
The SVM achieves a global accuracy of 99.7\%, where the misprediction occurs  between $m=2000$ and $m=8000$ which is the ``edge'' between the CPU and the XPU.
In all other intervals, the prediction is always correct.
The best accuracy is achieved with non-squared matrices, while square matrices give slightly lower accuracy.
Overall, this is a highly accurate predictor with a negligible runtime overhead of $`\leq 0.3msec$.

\subsection{Convolutions}
\begin{figure}[t]
\includegraphics[width=\linewidth]{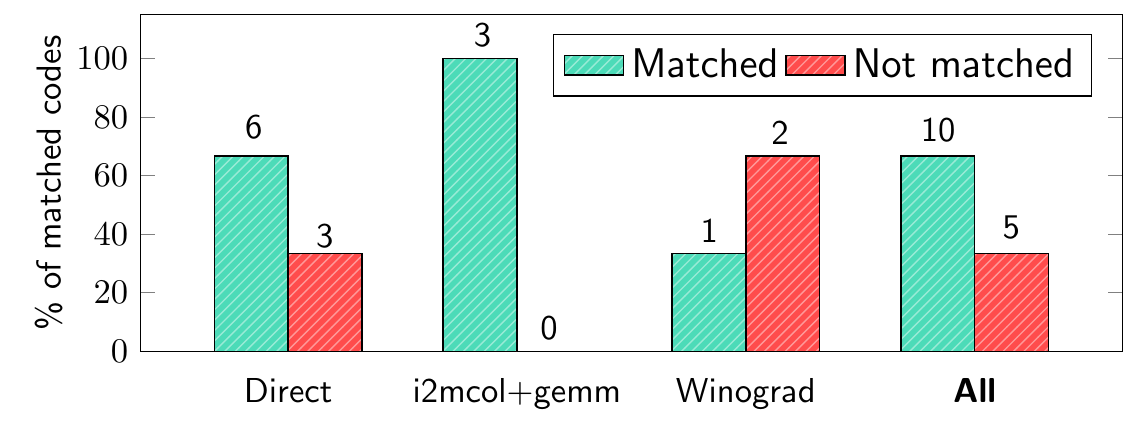}
\caption{Matched convolution codes by ATC.}
\label{fig:matched_conv}
\end{figure}

\begin{figure}[t]
\includegraphics[width=\linewidth]{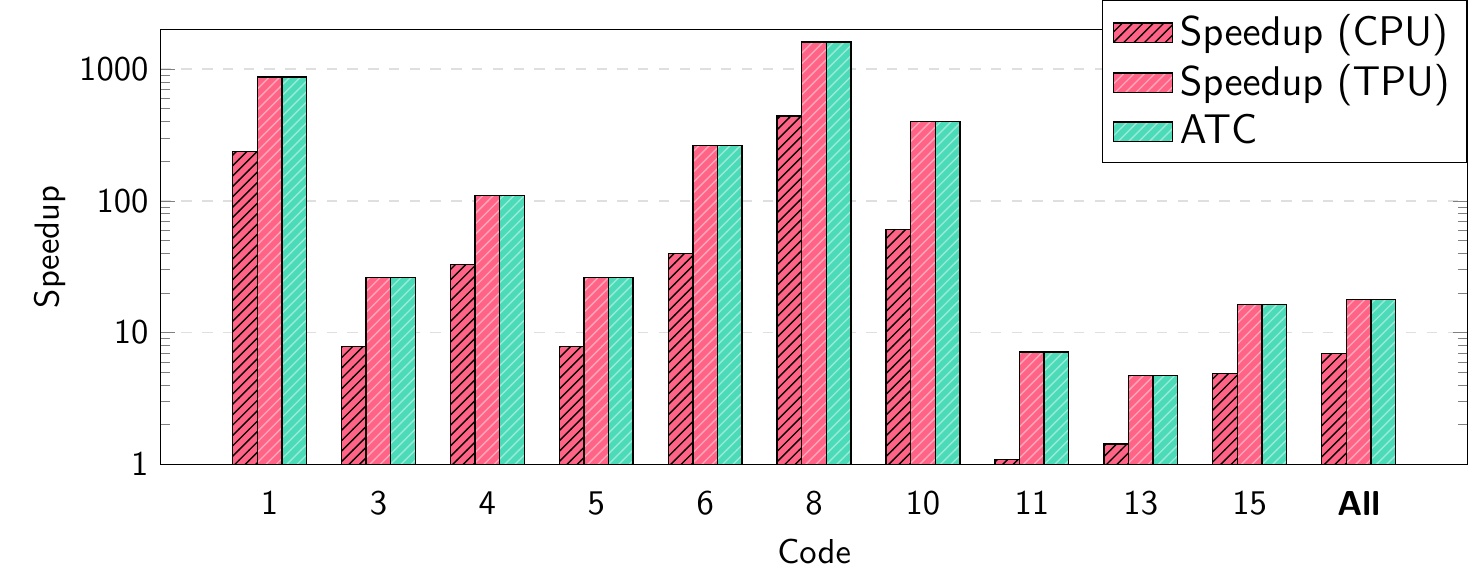}
\caption{ATC speedup in convolution programs with $h=w=224$, $kw=kh=11$, $c=3$, $k=96$ and $n=100$.}
\label{fig:speedups_conv}
\end{figure}

Our approach is generic and can be applied to other APIs other than GEMMs. 
As an example, we consider tensor convolutions which are a significant component of DNN workloads. 
While IDL, KernelFaRer, Polly and FACC* were unable to detect any of the convolutions, ATC detected 10 of the 15 convolutions as shown in Figure \ref{fig:matched_conv};
we were unable to match 5 due to the excessive number of candidates. 

Figure~\ref{fig:speedups_conv} shows the performance achieved by replacing with library code
for each of the programs we are able to accelerate. Across all codes, the  SVM predicts that the TPU accelerator outperforms the  CPU, giving an average 17.8x performance improvement across the programs.

\section{Related work}
\paragraph{Matching in Programs}
Matching high-level program structure has been used
to discover parallelism~\cite{di1996pap},
heterogenous offloading~\cite{andion2015compilation,Murray2011} and
many other core compiler tasks~\cite{MatchPD}.
Constraint languages make these tasks easier~\cite{idl,Blindell2018,MatchPD} but their
constraints are very sensitive to code structure~\cite{kernelfarer}.

For matrix multiplications in particular,
KernelFaRer~\cite{kernelfarer} provides a more robust approach,
detecting characteristics that define matrix multiplications.
Polyhedral analyses can also be used to target matrix multiplication
accelerators~\cite{Bhaskaracharya2020,Sun2021}, but both these techniques
fail to scale to the diversity of real code. 
FACC~\cite{facc} uses IO equivalence, which is robust to program structure, but only addresses the challenges of FFTs and does not scale to longer function signatures used for GEMM.
To support any accelerator type, the compiler should support multi-dimensional arrays, while FACC only supports 1D arrays.
Because in 1D arrays and FFTs the search space in matching the API parameters is small, FACC does not include anything to reduce it.
With more complex programs and domains, this limitation makes compiling programs intractable.

Mask~\cite{Samak2020} uses symbolic execution to prove equivalence, which does not work well for floating-point problems.
Fuzzy classification techniques based on code clone
detection~\cite{Lu2021,Su2016}, domain-classification~\cite{Uhrie2021}, pattern matching~\cite{collie2019type}, code embeddings~\cite{Alon2019,Allamanis2015,DeFreez2018} and
identifiers~\cite{Numata2016,Klainongsuang2019} can be used
to help compile to accelerators~\cite{facc}.  These
classification strategies are able to classify diverse
code structures, but do not provide a compilation strategy
for using an accelerator on their own.

A large class of techniques focus on migrating \textit{between}
APIs.  These techniques often use program synthesis~\cite{Collie2020},
NLP~\cite{Ni2021} and code embeddings~\cite{Nguyen2017,Phan2017}.
These techniques are unable to extract existing code into APIs.

\paragraph{Compiling for GEMM Accelerators}
Existing compilation strategies largely focus on \textit{lowering}
code from intrinsics to accelerators using rewrite rules~\cite{Steuwer2015a,Schlaak2022,Weng2021}
and synthesis techniques~\cite{Cowan2020}.

Existing approaches to extracting matrix multiplications~\cite{idl,kernelfarer} are brittle.
Synthesis-based techniques~\cite{Ahmad2019,Mendis2015,Angstadt2020a}
and rewriting-based techniques~\cite{chelini2021progressive,Steuwer2016}
have been developed to extract these DSLs that can then
be lowered: but they largely require flexible DSLs, rather than
APIs presented by hardware accelerators. 

\paragraph{Performance Prediction}
Predicting code the performance of hardware accelerators
is challenging, as the break-even point may depend on
many different arguments within a function's interface~\cite{logca}.
LogCA~\cite{logca} introduces static performance comparison
models for hardware accelerators and similar models have been applied
in offloading tasks~\cite{Yuan2020}.
Machine learning has often been applied in profitability settings,
such as  OpenCL Kernels~\cite{openclmapping,mergeorseparate} and
OpenMP~\cite{Mishra2020}.
Similar techniques have been applied to FPGAs,
by estimating power/performance~\cite{Fuhr2019} and tracking actual performance~\cite{Rigamonti2016}.

\section{Conclusions} \label{sec:conclusions}

This work presented ATC, a flexible domain-agnostic compiler that matches legacy linear algebra code to accelerators.
By using IO behavioral equivalence and smart search space reduction, we are able to match over 80\% of challenging real-world programs to accelerator APIs, significantly outperforming all alternative approaches.

Supporting new domains different from GEMM and convolution is easy because ATC focuses on behavior rather than code structure, which makes it very flexible and extensible.
Furthermore, to support other accelerators in GEMM or convolution, only the accelerator API is needed: ATC adapts to the new specification automatically.

Future work will examine  how to further reduce the search space using online learning and to expand the complexity of user code considered.
Longer-term, we wish to automatically target a range of accelerators  with diverse functionality, matching and transforming user code to maximize performance.

\begin{acks}
Grant TED2021-129221B-I00 funded by MCIN/AEI/10.13039/\\501100011033 and by the ``European Union NextGenerationEU/PRTR''.
\end{acks}

\bibliographystyle{ACM-Reference-Format}
\bibliography{refs}

\end{document}